\documentclass[reprint,
 amsmath,amssymb,
 aps,twocolumn,floatfix,prl
]{revtex4-1}

\usepackage[dvipsnames]{xcolor}
\usepackage{latexsym, color}
\usepackage{lipsum}
\usepackage[export]{adjustbox}
\usepackage{graphicx}
\usepackage{dcolumn}
\usepackage{bm}
\usepackage{amsmath,amsfonts,bm,graphicx,amssymb,color, amsthm, mathtools,amsbsy,setspace,sidecap}
\usepackage[left]{lineno}

\begin{document}

\preprint{APS/123-QED}

\title{CO$_2$-driven diffusiophoresis for removal of bacteria}

\author{Suin Shim$^{1}$}
\email{sshim@princeton.edu}
\author{Sepideh Khodaparast$^2$}
\author{Ching-Yao Lai$^3$}
\author{Jing Yan$^4$}
\author{Jesse T. Ault$^5$}
\author{Bhargav Rallabandi$^6$}
\author{Orest Shardt$^7$}

\author{Howard A. Stone$^{1,}$}%
\email{hastone@princeton.edu}
\affiliation{ \vspace{1ex}\\$^1$Department of Mechanical and Aerospace Engineering, Princeton University,
 Princeton, NJ 08544, USA \\
 $^2$School of Mechanical Engineering, University of Leeds, Leeds LS2 9JT, UK\\
 $^3$Lamont-Doherty Earth Observatory, Columbia University, Palisades, NY 10964, USA\\
 $^4$Department of Molecular, Cellular and Developmental Biology, Quantitative Biology Institute, Yale University, New Haven, CT, 06511, USA\\
 $^5$School of Engineering, Brown University, Providence, Rhode Island 02912, USA\\
 $^6$Department of Mechanical Engineering, University of California, Riverside, California 92521, USA\\
 $^7$Bernal Institute and School of Engineering, University of Limerick, Castletroy, Limerick, V94 T9PX, Ireland \looseness=-9
 \vspace{0.3ex}
}


\setlength{\skip\footins}{0.5cm}


\begin{abstract}
\vspace{1ex}
{We investigate CO$_2$-driven diffusiophoresis of colloidal particles and bacterial cells in a Hele-Shaw geometry. Combining experiments and a model, we understand the characteristic length and time scales of CO$_2$-driven diffusiophoresis in relation to system dimensions and CO$_2$ diffusivity. Directional migration of wild-type \textit{V. cholerae} and a mutant lacking flagella, as well as \textit{S. aureus} and \textit{P. aeruginosa}, near a dissolving CO$_2$ source shows that diffusiophoresis of bacteria is achieved independent of cell shape and Gram stain. Long-time experiments suggest possible applications for bacterial diffusiophoresis to cleaning systems or anti-biofouling surfaces.}

\begin{description}
\item[PACS numbers]
\end{description}
\end{abstract}

\maketitle



An aqueous suspension of charged particles in contact with a dissolving CO$_2$ source shows directional migration by diffusiophoresis \cite{shim2,shin}. Here we use a Hele-Shaw geometry with either a CO$_2$ bubble \cite{shim1,scb1, hsbprf} or a CO$_2$-pressurized chamber to investigate the transport of polystyrene particles and bacterial cells. Combining experiments and model calculations, we understand the characteristic length and time scales of CO$_2$-driven diffusiophoresis in relation to the system dimensions and CO$_2$ diffusivity. We then study migration of wild-type \textit{Vibrio cholerae} and a mutant lacking flagella ($\Delta$\textit{flaA}) near a dissolving CO$_2$ source, showing that the directional motion is \textit{diffusiophoresis}, not CO$_2$-driven \textit{chemotaxis}. Also, we demonstrate diffusiophoresis of \textit{Staphylococcus aureus} \cite{SAmKO} and \textit{Pseudomonas aeruginosa} \cite{PA14}, showing that diffusiophoresis driven by CO$_2$ dissolution occurs for both Gram-positive and Gram-negative bacteria, independent of shape. We further demonstrate that diffusiophoretic removal of \textit{S. aureus} reduces cell adhesion to a surface, and that the removal of \textit{P. aeruginosa} lasts $\geq 11$ hr after CO$_2$ is turned off. CO$_2$-driven diffusiophoresis can prevent surface contamination or infection by reducing the population of cells approaching an interface, and the mechanism can be applied to liquid cleaning systems and anti-biofouling surfaces.

\begin{figure*}[t!]
\centering
\includegraphics[width=6.3in]{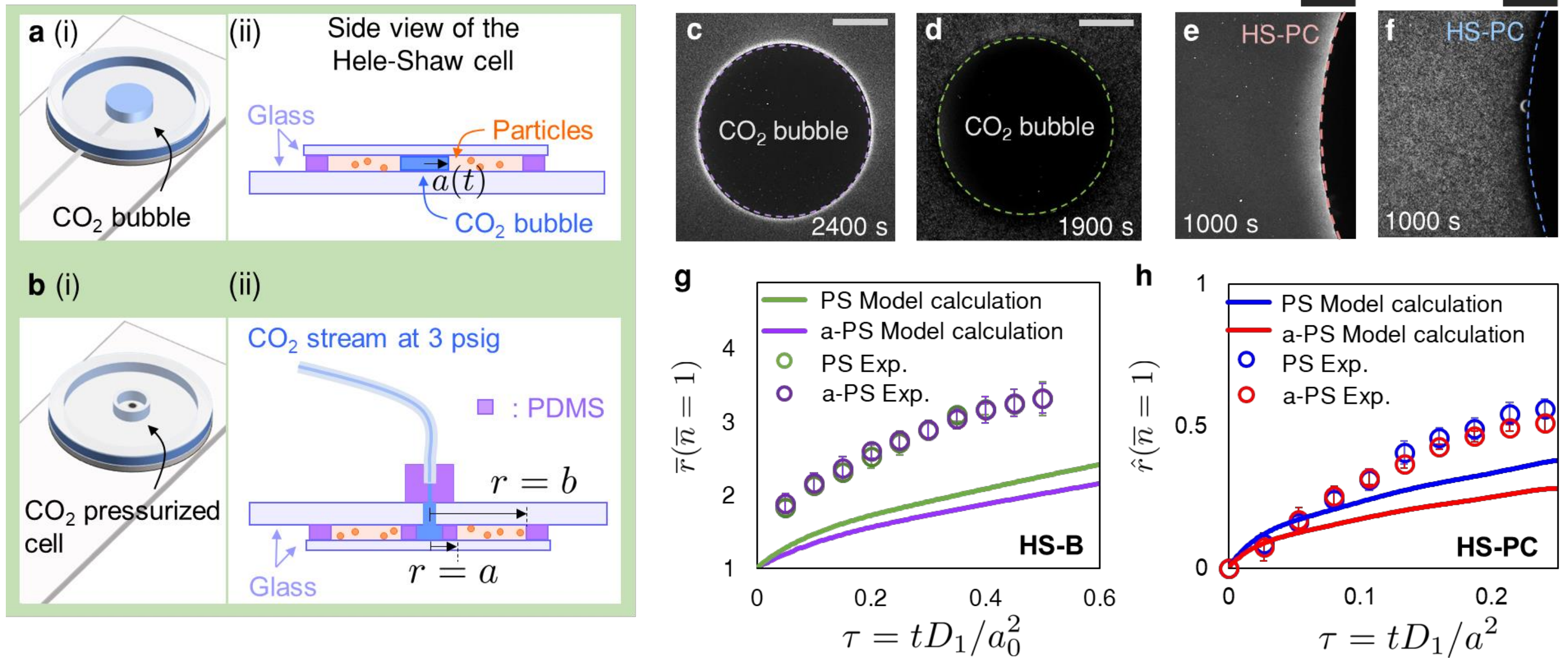}
\vspace{-1.2ex}
\caption{\label{fig1} {CO$_2$-driven diffusiophoresis of colloidal particles. (a,b) Schematics of experimental setup for (a) HS-B and (b) HS-PC. See SI for details. (c,d) Charged particles near a dissolving CO$_2$ bubble (HS-B). Distribution of (c) amine-modified polystyrene (a-PS) particles and (d) polystyrene (PS) particles show, respectively, local accumulation and exclusion of charged particles by diffusiophoresis. Bright dots indicate particles. (e,f) Charged particles near the CO$_2$ source in HS-PC. Distribution of (e) a-PS and (f) PS particles near the fixed CO$_2$ source show local accumulation and exclusion. (g,h) Comparison between experimental measurements and model calculations of the macroscopic growth of the accumulation and exclusion zones. (g) Measured and calculated values of $\bar{r}(\bar{n}=1)$ are plotted versus $\tau$ for HS-B. (h) Measured and calculated values of $\hat{r}(\bar{n}=1)$ are plotted versus $\tau$ for HS-PC. (g,h) No fitting parameter is used. (c-f) Scale bars are 500 $\mu$m.}}
\vspace{-2.3ex}
\end{figure*}

When an aqueous suspension of charged colloidal particles is exposed to dissolving CO$_2$, positively (negatively) charged particles migrate toward (away from) the CO$_2$ source by diffusiophoresis \cite{shim2,shin}; the fast diffusing H$^+$ relative to HCO$_3^-$ from the dissolution of H$_2$CO$_3$ drives the transport. We investigate the phenomenon using a Hele-Shaw geometry (Fig. \ref{fig1}; a circular cell with radius $b=11$ mm and height $h =500~\mu$m). Here, diffusiophoresis near a CO$_2$ source is documented experimentally and calculated (Supplementary Information; SI) for two configurations -- a dissolving CO$_2$ bubble (Fig. \ref{fig1}(a); we call this system HS-B) and a CO$_2$-pressurized chamber (Fig. \ref{fig1}(b); HS-PC) -- to examine both moving and fixed boundaries.

In HS-B, a CO$_2$ bubble with radius $a(t)$ and an initial radius $a_0$ dissolves at a typical speed $\frac{da}{dt}\approx \frac{D_1}{a_0}\approx O(0.1$-$1)~\mu$m/s until the gas exchange reaches steady state \cite{shim2,shim1,scb1,hsbprf}, where $D_1$ is the diffusivity of CO$_2$ in water. A bubble reaches its steady state within ${\tau=\frac{t}{a_0^2/D_1}\approx1}$ (SI). The typical diffusiophoretic velocity of nearby suspended particles $u_p=\Gamma_p \nabla \ln c_i$ scales as $u_p \approx \frac{\Gamma_p}{a_0} \approx O(0.1$-$1)~\mu$m/s, where $\Gamma_p$ and $c_i$ are the diffusiophoretic mobility of particles and concentration of ions, respectively (SI, video 1). The relative motion of particles and the interfaces creates a charge-dependent particle distribution for both HS-B and HS-PC (Fig. \ref{fig1}(c-f)).

Locally, in the vicinity of the interface, amine-modified polystyrene (a-PS, positively charged, diameter $=$ 1 $\mu$m) particles accumulate and form a high particle-density region, whereas polystyrene (PS, negatively charged, diameter $=$ 1 $\mu$m) particles create an exclusion zone (EZ; Fig. \ref{fig1}(c-f)), where the particle concentration is small. Growth of the local EZ in HS-PC (SI Fig. S2) is proportional to $\sqrt{t}$, similar to EZ formation near an ion-exchange membrane \cite{ez1}.

Particle accumulation and exclusion also occur on the length scale $\ell \approx a_0$ in both systems (Videos 2,3). In the model we define the boundaries of macroscopic accumulation and exclusion as the radial distance where the nondimensional particle concentration $\bar{n}=n/n_0= 1$ ($n(r,t)$ is the particle concentration and $n_0=n(r,0)$; SI). The nondimensional radial positions are defined as $\bar{r}=r/a_0$ for HS-B, and ${\hat{r}=\frac{r - a}{b-a}}$ for HS-PC. Such boundaries are determined analogously in the experiments (SI) and plotted versus $\tau$ in Fig. \ref{fig1}(g,h). For HS-B, the boundaries grow faster in experiments due to the initial rapid generation of the bubble, which is not included in the model. Bubble generation introduces fast interface growth, which enhances CO$_2$ dissolution at the early times and causes faster diffusiophoreis. The particle dynamics show better agreement in HS-PC. Without the initial growth in the measured boundaries in HS-B, we obtain similar trends of the particle dynamics between HS-B and HS-PC (SI). 

The macroscopic boundaries increase up to almost half of the radius of the Hele-Shaw cell ($\approx 0.5b$) within $\tau = 0.2$ in HS-PC. In HS-B, as noted from the time evolution of the radius (SI), there is a velocity contribution from the shrinking bubble ($da/dt$) that affects the particle distribution, and this effect lasts up to $\tau \approx 1$. 


Our understanding of the typical length and time scales of CO$_2$-driven diffusiophoresis in a Hele-Shaw cell motivated us to extend our investigations to a broader range of particles. Past studies have reported on the use of diffusiphoresis to achieve migration of living cells \cite{cell1,cell2}. For example, the goals of particle manipulation can be to clean a region of liquid, achieve antifouling surfaces, or prevent infection in biological systems. Two previous studies report EZ formation in bacterial suspensions in contact with an ion-exchange membrane (Nafion) \cite{bac1, bac2} and discuss possible cleaning applications. 

As an initial step for demonstrating and investigating diffusiophoresis of bacterial cells by CO$_2$ dissolution, we chose two types of \textit{V. cholerae} cells -- wild-type (WT) and a mutant lacking flagella ($\Delta$\textit{flaA}), both of which are tagged by mKO (monomeric Kusabira Orange), a bright fluorescent protein \cite{mko}. We first confirm the diffusiophoretic contribution to the cell migration in the presence of a dissolving CO$_2$ source in a Hele-Shaw geometry. Then, using PIV, we measure the velocities of the bacterial cells that move along the ion concentration gradient. 

\textit{V. cholerae} is Gram-negative, comma-shaped (length $\approx 2$-3$~\mu$m, diameter $\approx 1~\mu$m), and single flagellated. The net surface charge of \textit{V. cholerae} (as well as other bacteria) is negative \cite{vccharge1, vccharge2}, so the cells are expected to migrate away from a CO$_2$ source by diffusiophoresis (Video 4). We prepared a bacterial solution by diluting the growth suspension (see SI for Methods) to 10$\%$ M9 minimal salt solution. No nutrient is provided so no growth and division occur on the time scale of the experiment. Using low salt concentration helps to exclude effects of coupled ion fluxes on the diffusiophoresis of bacteria \cite{FSS}. Similar to the particle experiments, we fill the Hele-Shaw cell with bacterial suspension, and introduce either a CO$_2$ bubble or pressurize CO$_2$ in the inner chamber. Fluorescence intensities near the CO$_2$ source for both HS-B and HS-PC systems are measured (Fig. S10), and the intensity change shows that the cell number near the dissolving CO$_2$ source decreased significantly over time. 
%

\vspace{-0.3ex}

\begin{figure}[t!]
\centering
\includegraphics[width=3.5in]{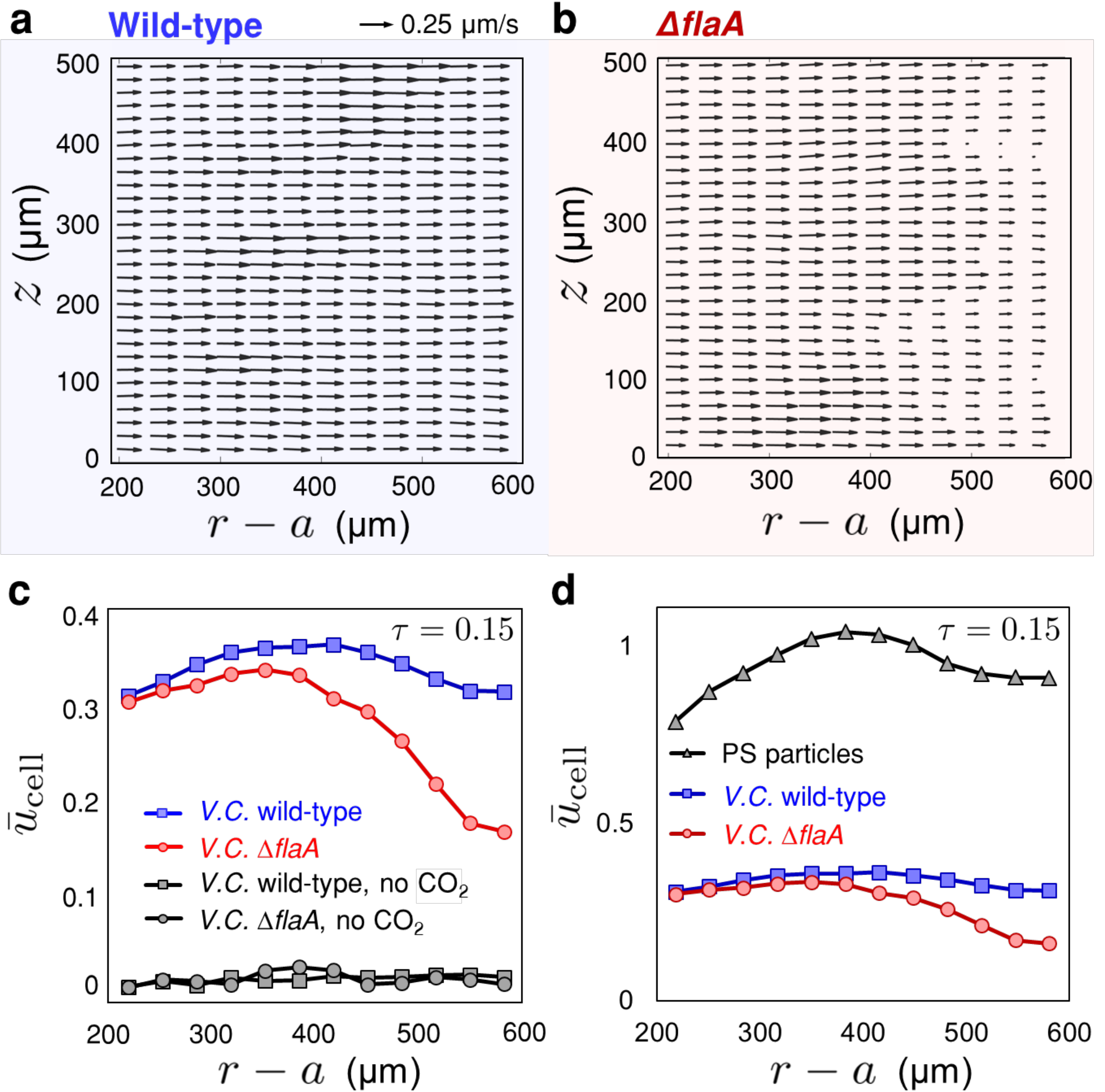}
\vspace{-3.6ex}
\caption{\label{fig2-2} {Velocity measurements for CO$_2$-driven diffusiophoresis of \textit{V. cholerae}. (a,b) PIV for \textit{V. cholerae} cells in the HS-PC experiments. Velocity vectors plotted versus position ($r-a$, $z$). Motion of (a) wild-type and (b) $\Delta$\textit{flaA} cells at $t \approx$ 10 minutes. The directional migration of cells is described by aligned velocity vectors in the radially outward direction. (c) Nondimensional $z$-averaged velocities obtained from (a,b) and control experiments without CO$_2$ at $\tau=0.15$ ($\approx 10$ minutes) plotted versus $r-a$. (d) Nondimensional $z$-averaged velocities of PS particles, WT and $\Delta$\textit{flaA} cells obtained at $\tau=0.15$ plotted versus $r-a$. } }
\vspace{-2.3ex}
\end{figure}

Particle image velocimetry (PIV) near the fixed boundary measures the diffusiophoretic velocity of the cells by a dissolving CO$_2$ source. We plotted the velocity vectors versus position in Fig. \ref{fig2-2}(a,b), where the origin of the $z$-axis is at the bottom left corner. After the CO$_2$ valve is opened at $\tau=0$, both strains of \textit{V. cholerae} migrate radially outward (Fig. \ref{fig2-2}(a,b)). The radial alignment of the velocity vectors confirms that both motile and immotile \textit{V. cholerae} cells move along the CO$_2$-generated ion concentration gradient. In Fig. \ref{fig2-2}(c), nondimensional $z$-averaged velocities ($\bar{u}_{\text{cell}} = u_{\text{cell}}/(D_1/a)$; $u_{\text{cell}}$ is the $z$-average of measured velocity) of the cells at $\tau=0.15$ with and without dissolving CO$_2$ are plotted. Our observation that both motile and immotile cells exhibit directional migration with similar velocities shows that the motion is not a chemotactic effect. We also compare the typical velocity scales of the cells and the PS particles in Fig. \ref{fig2-2}(d). The diffusiophoretic velocity of the bacterial cells is smaller than that of the PS particles, and as a first rationalization, this is due to the smaller diffusiophoretic mobility of the cells. Our comparison suggests that the \textit{V. cholerae} cells have three to four times smaller mobility compared to the PS particles, since the diffusiophoretic velocity scales as $u_p \approx \frac{\Gamma_p}{a}$.

\begin{figure*}[t!]
\centering
\includegraphics[width=6.5in]{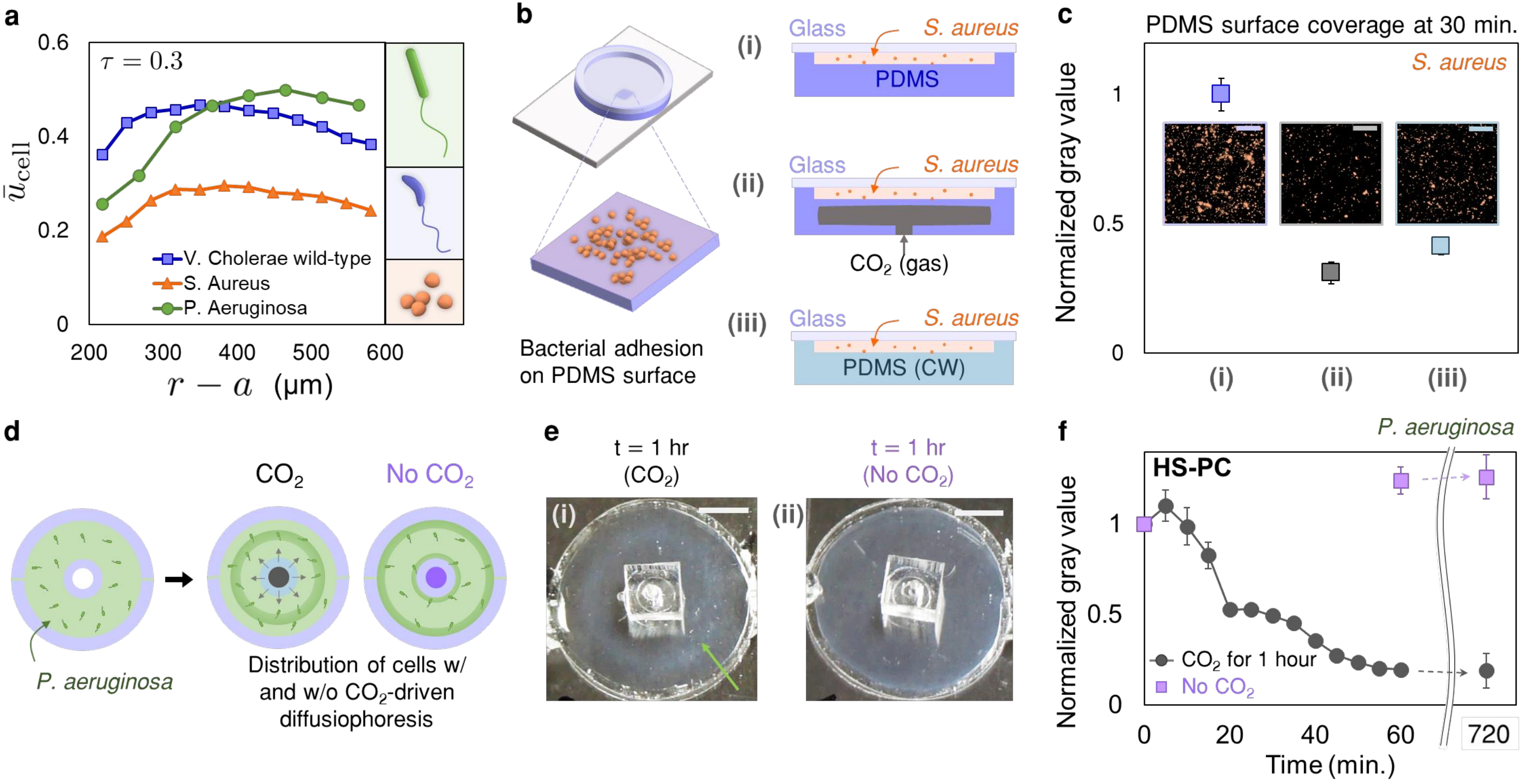}
\vspace{-3.33ex}
\caption{\label{fig3-2} {CO$_2$-driven diffusiophoresis of \textit{S. aureus} and \textit{P. aeruginosa}. (a) Velocity measurements of bacterial cells near the CO$_2$ source (fixed boundary configuration) at $\tau=0.3$ ($t \approx 20$ min). Right panels: schematics showing cell shape. (b) Schematics of adhesion experiments for \textit{S. aureus} cells on PDMS surfaces. We tested three scenarios: (b-i) plain PDMS substrate in ambient air, (b-ii) pressurized CO$_2$ gas under the PDMS substrate, and (b-iii) PDMS substrate that is saturated with carbonated water (CW). (c) Intensity measurements for cell coverage on PDMS substrate at $t=30$ min. The gray values are normalized by that of a corresponding no-CO$_2$ experiment. Inset images show attachment of \textit{S. aureus} cells on three different PDMS surfaces. (d-f) Long time effect of diffusiophoresis: diffusiophoresis of \textit{P. aeruginosa} cells in the fixed boundary system. (d) Schematic of two control experiments in the fixed boundary configuration. In the presence of a finite-time CO$_2$ source, cells move radially outward and form an accumulation front (ring structure) in the chamber. On the other hand, when the CO$_2$ source is replaced by an air source, cells gradually concentrate toward both PDMS walls. (e-i,ii) Images showing the Hele-Shaw chamber at $t=1$ hr. (f) Fluorescent intensity measurements near the CO$_2$ source for two experiments in (d,e). Accumulation and exclusion of bacterial cells near the inner PDMS wall is maintained up to 12 hours, proving long-term effect of diffusiophoresis. Scale bars are (c) 50 $\mu$m, and (e) 5 mm.}}

\end{figure*}

To highlight the generality of the phenomenon, two more bacteria were examined -- \textit{S. aureus} (mKO labeled, Gram-positive, spherical and immotile) \cite{SAmKO} and \textit{P. aeruginosa} (mCherry labeled, Gram-negative, rod shaped and motile) \cite{PA14} for their diffusiophoretic response to dissolving CO$_2$ (Fig. \ref{fig3-2}). For the HS-PC system, the diffusiophoretic velocities at $\tau=0.3$ are plotted versus ${r-a}$ in Fig. \ref{fig3-2}(a). We note that \textit{S. aureus} is slower compared to the other two bacteria. Both \textit{S. aureus} and \textit{P. aeruginosa} have surface zeta potentials $\zeta \approx -30$ mV \cite{SAzeta1,SAzeta2,PAzeta1}, so the velocity difference is unexpected given that electrophoresis makes the dominant contribution to CO$_2$-driven diffusiophoresis, where $\Gamma_p$ is a function of $\zeta$ \cite{prieve1984} (SI). One feature of \textit{S. aureus} is that surface adhesive proteins \cite{sticky} make the cells easily form clusters, which can contribute to the change in the diffusiophoretic velocity. Also, different diffusiophoretic velocities may arise from different shapes of the cells (Fig. \ref{fig3-2}(a)). \textit{S. aureus} is approximately spherical with a diameter $\approx 1~\mu$m, while \textit{P. aeruginosa} is $\approx 1$-5 $\mu$m long and $\approx 0.5$-$1~\mu$m in diameter. Assuming similar $\zeta$ for the three bacteria and the largest aspect ratio for \textit{P. aeruginosa}, we obtain an aspect-ratio dependence of the diffusiophoretic velocities of the cells \cite{shape}. Our results also show that the surface zeta potential is not the only parameter for determining the diffusiophoretic velocity of bacterial cells.

In order to move toward applications for diffusiophoretic bacterial removal using CO$_2$, we first quantify adhesion of \textit{S. aureus} cells to surfaces under different CO$_2$ dissolution conditions. Fig. \ref{fig3-2}(b) illustrates three conditions of PDMS substrates (b-i) without and (b-ii,iii) with CO$_2$ sources. CO$_2$ is introduced either by pressurizing CO$_2$ below a PDMS membrane or by saturating PDMS with carbonated water (CW; see SI for Methods). Then the surface coverage 30 minutes after injection of a \textit{S. aureus} suspension into the chamber above the PDMS substrate was measured by the fluorescent intensity (Fig. \ref{fig3-2}(c)). We observe that the surface attachment of \textit{S. aureus} is significantly decreased in the presence of CO$_2$, and this is evidence that the CO$_2$-generated ion concentration gradient removed the cells from the vicinity of the substrate, resulting in reduced surface contamination. Below, with a set of experiments with \textit{P. aeruginosa}, we demonstrate that the bacterial removal lasts $\geq 11$ hours.

In many discussions of diffusiophoresis, the focus is often on boosting migration of micron-sized particles. This is a clear advantage of the phenomenon, owing to $\Gamma_p \gg D_p$ ($D_p \approx 10^{-13}$ m$^2$/s is the Stokes-Einstein diffusivity of a micron-sized particle). However, smaller particle diffusivity compared to the diffusiophoretic mobility, can also mean that, after eliminating the gradient, the time required for particles to recover their original distribution is long ($\approx 1000$ hr for 1 $\mu$m particles to move 1 mm by diffusion; SI). 

\textit{P. aeruginosa} is known for surviving in dilute media \cite{PAsurv} for more than 10 days so it is suitable for long-time diffusiophoretic experiments. We performed HS-PC experiments with and without 1 hr of CO$_2$ dissolution in \textit{P. aeruginosa} suspension (Fig. \ref{fig3-2}(d,e)). We predict and observe that, by CO$_2$ diffusiophoresis, bacterial cells move away from the inner wall, whereas without any CO$_2$ source, the cells concentrate near both inner and outer PDMS walls where there is an air source. The CO$_2$ valve was open only for 1 hr, but the result of diffusiophoresis lasted longer than 12 hours (Fig. \ref{fig3-2}(f)). The distribution of the cells at $t=12$ hr are presented in the SI.

Finally, we discuss the diffusiophoresis of \textit{motile} bacteria since it is not identical to that of polystyrene particles or immotile cells. Both \textit{V. cholerae} and \textit{P. aeruginosa} are single flagellated organisms and exhibit run-reverse patterns \cite{bacSpeed}. The effective diffusivity of motile bacteria with typical translational speed $v_t$ and reverse time $t_r$ can be estimated as $D_{\text{eff}}\approx v_t^2 t_r \approx O(100)~\mu$m$^2$/s (SI). It is observed (Video 5) that the flow of cells under ion concentration gradient is a slow advection with an estimated P\'eclet number $Pe = \frac{ u_p \ell_{\text{cell}}}{D_{\text{eff}}} \approx 10^{-3}$-$10^{-2}$. Cells are observed to swim randomly with their characteristic velocity $\approx 30$-$50~\mu$m (SI), with a slow drift (radially outward) due to the diffusiophoretic contribution (Video 5).

In this paper, we present proof of diffusiophoretic migration of different types of bacteria under a concentration gradient of CO$_2$, and discuss possible applications of CO$_2$-driven diffusiophoresis to prevent contamination. For example, delaying biofilm formation can improve the anti-biofouling properties of surfaces. Currently we are working to realize the mechanism at various salt concentrations to broaden the understanding to physiological or higher salinity conditions. Moreover, understanding the characteristic scales and flow stucture near the CO$_2$ source is crucial for the next steps of CO$_2$-driven diffusiophoresis for mitigating bacterial growth on, or bacterial removal from, surfaces.

\section*{Acknowledgements}
We thank the Bassler Lab for providing the $V.~cholerae$ strains (JY019 and JY238) for the current study. S.S. thanks Minyoung Kim and Christina Kurzthaler for valuable discussions. S.S. and H.A.S. acknowledge the NSF for support via CBET-1702693. S.K. thanks L’Oréal-UNESCO UK and Ireland for support via the FWIS 2019 fellowship.

\section*{Contributions}

S.S. and H.A.S. conceived the project. S.S. designed and performed all experiments.
S.K. conducted PIV. S.S., C.Y.L., J.T.A. conducted numerical calculations. J.Y. constructed the \textit{V. cholerae} strains. S.S., B.R., O.S., and H.A.S. set up the theoretical model. All authors contributed to data analysis and writing the paper. 


\end{document}